\documentclass[11pt]{article}
\begin{document}

\newfam\msbfam
\batchmode\font\twelvemsb=msbm10 scaled\magstep1 \errorstopmode
\ifx\twelvemsb\nullfont\def\Bbb{\bf}
        \font\fourteenbbb=cmb10 at 14pt
	\font\eightbbb=cmb10 at 8pt
	\message{Blackboard bold not available. Replacing with boldface.}
\else   \catcode`\@=11
        \font\tenmsb=msbm10 \font\sevenmsb=msbm7 \font\fivemsb=msbm5
        \textfont\msbfam=\twelvemsb
        \scriptfont\msbfam=\tenmsb \scriptscriptfont\msbfam=\sevenmsb
        \def\Bbb{\relax\expandafter\Bbb@}
        \def\Bbb@#1{{\Bbb@@{#1}}}
        \def\Bbb@@#1{\fam\msbfam\relax#1}
        \catcode`\@=\active
	\font\fourteenbbb=msbm10 at 14pt
	\font\eightbbb=msbm8
\fi
\catcode`\@=11
\def\Z {{\Bbb Z}}
\def\R {{\Bbb R}}
\def\E {{\Bbb E}}
\newfam\scrfam
\batchmode\font\twelvescr=rsfs10 at 12pt \errorstopmode
\ifx\twelvescr\nullfont
        \message{rsfs script font not available. Replacing with calligraphic.}
        \def\scr{\cal}
\else   \font\tenscr=rsfs10 
        \font\sevenscr=rsfs7
        \skewchar\twelvescr='177 \skewchar\tenscr='177 \skewchar\sevenscr='177
        \textfont\scrfam=\twelvescr \scriptfont\scrfam=\tenscr
        \scriptscriptfont\scrfam=\sevenscr
        \def\scr{\fam\scrfam}
        \def\cal{\scr}
\fi
\def\unit{\hbox to 3.3pt{\hskip1.3pt \vrule height 7pt width .4pt \hskip.7pt
\vrule height 7.85pt width .4pt \kern-2.4pt
\hrulefill \kern-3pt
\raise 4pt\hbox{\char'40}}}
\def\II{{\unit}}
\def\cM {{\cal{M}}}
\def\half{{\textstyle {1 \over 2}}}
\newcommand{\od}{\widetilde{\rm OD}}
\def    \beq    {\begin{equation}} \def \eeq    {\end{equation}}
\def    \bea    {\begin{eqnarray}} \def \eea    {\end{eqnarray}}
\def\la{\label} \newcommand{\eq}[1]{(ref{#1})}
\def    \lf     {\left (} \def  \rt     {\right )}
\def    \a      {\alpha} \def   \lm     {\lambda}
\def    \D      {\Delta} \def   \r      {\rho}
\def    \th     {\theta} \def   \rg     {\sqrt{g}} \def \Slash  {\, /
\! \! \! \!}  \def      \comma  {\; , \; \;} \def       \pl
{\partial} \def         \del    {\nabla}
\newcommand{\mx}[4]{\left#1\begin{array}{#2}#3\end{array}\right#4}
\newcommand{\Dpp}{\Delta + \nu}
\newcommand{\Dmm}{\Delta - \nu}
\newcommand{\Dp}{\Delta_+}
\newcommand{\Dm}{\Delta_-}
\newcommand{\Ds}{\left(\Delta^2-\nu^2\right)}
\newcommand{\Pp}{\Pi_+}
\newcommand{\Pm}{\Pi_-}
\newcommand{\lp}{\ell_p}
\newcommand{\ie}{{\em i.e., }}
\newcommand{\eg}{{\em e.g., }}
\newcommand\sss{\scriptscriptstyle}
\newcommand\scs{\scriptstyle}
\newcommand{\bc}{\begin{center}}
\newcommand{\ec}{\end{center}}
\newcommand{\nz}{\normalsize}
\newcommand\nn{\nonumber}
\frenchspacing
\begin{titlepage}
\begin{flushleft}
       \hfill                      {\tt hep-th/0201178}\\
       \hfill                        G\"oteborg-ITP-preprint\\
\end{flushleft}
\vspace*{1.8mm}
\begin{center}
\LARGE {\bf Light-like noncommutativity and duality from \\ 
open strings/branes}\\
\normalsize
\vspace*{5mm}
{\Large
Henric Larsson}\\
\vspace*{2mm}
{\small Institute of Theoretical  Physics\\
G\"{o}teborg University and Chalmers University of Technology\\
SE-412 96 G\"{o}teborg, Sweden \\
E-mail: solo@fy.chalmers.se}\\
\vspace*{3mm}
\end{center}
\Large \begin{center}
{\bf Abstract}
\end{center}
\normalsize
In this paper we perform some non-trivial tests for the recently obtained 
open membrane/D-brane metrics and `generalized' noncommutativity 
parameters using \mbox{D$p$/NS5/M5-branes} which have been deformed by 
light-like fields. The results obtained give further evidence that these 
open \mbox{membrane/D-brane} metrics and `generalized' noncommutativity 
parameters are correct. Further, we 
use the open brane data and supergravity duals to obtain more information 
about non-gravitational theories with light-like noncommutativity, or 
`generalized' light-like noncommutativity. In particular, we investigate
 various duality
 relations (strong coupling limits). In the light-like case we also comment on 
the relation between open membrane data (open membrane metric etc.) in six 
dimensions and open string data in five dimensions. Finally, we 
investigate the strong coupling limit (high energy limit) of five 
dimensional NCYM with $\Theta^{12}=\Theta^{34}$. In particular, we find that 
this NCYM theory can be UV completed by a DLCQ compactification of M-theory.

\end{titlepage}

\section{Introduction}
During the last couple of years there have been a huge interest in 
non-gravitational theories 
which have a noncommutative structure. The first so called noncommutative 
theory to be obtained was noncommutative super Yang-Mills (NCYM), with 
space-space \cite{con}-\cite{Delle} or light-like \cite{gomis}-\cite{paolo} 
noncommutativity. This was 
 later followed by noncommutative open string theory 
(NCOS) \cite{gopa}-\cite{mikkel}, as well as various world volume theories 
(of deformed D-branes, NS5-branes and M5-branes) 
containing light open M2-branes or open D-branes, 
OM/OD$p$/$\widetilde{{\rm OD}q}$ \cite{GMSS}-\cite{gui,NS5}. 

It is well known that the open string metric, coupling constant and 
noncommutativity parameter \cite{sw}, give important 
information for both NCYM and NCOS (e.g. in the mass shell condition for 
NCOS and in the definition of the Yang-Mills coupling constant for NCYM). These
 open string quantities are well defined and obtained from the open 
string two-point function \cite{sw} (metric and noncommutativity parameter) 
and the Dirac-Born-Infeld action (coupling constant, see e.g. \cite{sw}). 
 For OM-theory it was conjectured in \cite{us2} that there is an 
open membrane (M2) metric which is relevant for this theory. Later in 
\cite{janpieter,openm,ericjanpieter}, the full open membrane metric was 
obtained, including the conformal factor, using different methods. 
Unfortunately we still lack a microscopic understanding of the open membrane
metric. In \cite{openm,ericjanpieter}, the full structure of a
conjectured open membrane generalized noncommutativity parameter was also 
obtained, again using different methods. This object is expected to be  
the open membrane analog of the open string noncommutative parameter, 
see \cite{openm,ericjanpieter,others} where the relation to the open string 
noncommutativity parameter is derived. In paper \cite{openm} 
expressions for open D$q$-brane metrics and generalized noncommutativity 
parameters were also obtained. These are expected to be important for the 
open D-brane theories ODp \cite{GMSS,harmark2,NS5,obers1,lu,soloper,od5} 
(NS5-brane world volume theories containing light open D$p$-branes) and  
$\widetilde{\rm ODq}$ \cite{lu,soloper} (D$(q+2)$-brane world volume theories 
containing light open D$q$-branes).

In this paper we will further investigate these open brane (D$q$/M2) 
metrics for the special case where we have turned on a light-like field 
on a D$p$/NS5/M5-brane. In particular, we show that if one insert 
a supergravity solution, which corresponds to a D$p$/NS5/M5-brane with 
a light-like field turned on, in these open metrics, one obtains a 
deformation independent \cite{Berman} open brane metric\footnote{This means 
that the 
result is independent of the deformation parameter $\theta$ in the solution, 
see section 2 and \cite{intr,Delle,openm}. This result is important, since the 
assumption of deformation independence played an important role in the 
derivation of the open membrane/D-brane metrics in \cite{openm}.}. We also 
show that one obtains the
 expected results after taking the near horizon limit (e.g. fixed generalized 
noncommutativity parameter in all cases). For a 
deformation with a magnetic or electric RR field, deformation independence 
 and fixed generalized noncommutativity, was derived in \cite{openm}. Further, 
 we find here that the 
open membrane metric and generalized noncommutativity parameter in six 
dimensions can be reduced to the open string metric, coupling constant and 
noncommutativity parameter in five dimension, in the case when the 
M-theory three form $A$ is light-like (light-like $A\rightarrow$ light-like 
NS-NS $B$, where $B$ is dual to a light-like RR three form $C$). In 
\cite{openm,ericjanpieter}, it was demonstrated
that the reduction for a general three form only works for the `electric' 
rank 2 case, see \cite{openm,ericjanpieter} for details.
However, the light-like case was excluded from the 
investigation. Further, we investigate 
field theories and Little String theories with light-light noncommutativity 
\cite{gomis,NS5}, specially 
their duality relations. We also include
 a discussion about the strong coupling limit of five dimensional NCYM with 
$\Theta^{12}=\Theta^{34}\neq 0$, which is given by a DLCQ compactification of 
M-theory in the presence of a `parallel' 
M5-brane (i.e., one of the M5-brane directions is in the light-like 
compactified direction) and two transverse M2-branes.

The paper is organized as follows: Section 2 contains the relevant 
supergravity duals, as well as the open F1/D$q$/M2 data, that are needed 
later. 
In section 3 we discuss various issues about field theories and 
Little String theories with light-like
`ordinary' or `generalized' noncommutativity, as well as how open string and 
open membrane data are related in five and six dimensions. 
In section 4 we investigate the strong coupling limit of NCYM with 
$\Theta^{12}=\Theta^{34}\neq 0$ in five dimensions. 
There are also two appendices included, where 
we have collected our conventions regarding T-duality, S-duality and 
lift from type IIA to eleven dimensions, as well as a discussion 
on the relation between the `supergravity dual' and the `flat space 
scaling' approaches to noncommutative theories.   

\section{Supergravity duals and open brane data}

In this section, we obtain the supergravity duals of all 
non-gravitational theories with a one parameter light-like noncommutativity, 
in dimensions $d\leq 6$\footnote{These supergravity duals have 
earlier been obtained in \cite{NS5} (for the deformed 
D$p$-brane cases, see also \cite{cai6}), using a different solution generating 
technique. We will here obtain equivalent solutions, using the O($p+1,p+1$) 
method \cite{Berman,Delle,solo1}, see below.}.
 The relevant open string/open D-brane data are also computed and shown to be 
deformation independent. 

The supergravity duals are obtained by starting with the relevant 
supergravity solution, corresponding to a bound state (stack of $N$ branes at 
$r=0$ with different fields turned on), followed by taking 
a near horizon limit (involves $\alpha'\rightarrow 0$, see below). The 
so obtained supergravity solution is called a supergravity dual. This 
supergravity dual is then probed by a probe brane. By taking this probe brane 
infinitely far away from the stack (i.e., $\tilde{r}\rightarrow\infty$), the 
U(1) degrees of freedom on the probe brane will decouple from gravity, and the 
physics of the probe brane
can be described by a non-gravitational theory with noncommutativity, see e.g.
 \cite{Berman,openm,cai4} for further clarifications. If e.g. 
the probe brane is 
a D$p$-brane, then open strings can end on it. Therefore, the open string 
metric, coupling constant and noncommutativity parameter give important 
information of the physics on this probe brane. For example, as we will see 
below, in 
the case of a D$p$-brane with a light-like $B$-field turned on, we obtain, 
using the supergravity dual and open string data, that the physics on the 
single probe brane is governed by a U(1) ($p+1$)-dimensional NCYM theory, with 
light-like noncommutativity $\Theta^{-1}\neq 0$\footnote{For a stack of $k$ 
probe D$p$-branes on top of each other, we instead have a U(k) 
theory, where $k<<N$, in order for the probe branes to have a 
negligible effect on the background. In this paper we usually set $k=1$.}. 

We note here, that this supergravity set up is equivalent to instead starting 
with flat 
space and then taking a decoupling limit a la e.g. \cite{sw,GMSS} in order to 
decouple gravity. In appendix B there is a further discussion on these two 
approaches and their equivalence. The main reason why we 
use the supergravity dual approach is because it gives us the opportunity 
to check that the recently derived open M2/D$q$-brane metrics 
\cite{janpieter,openm} are deformation independent \cite{Berman,openm}, in the
case of light-like deformations. 
That the open M2/D$q$-brane metrics are deformation independent, implies that 
if one insert a deformed brane solution (see below) in the 
relevant open M2/D$q$-brane metrics, then the result is not dependent 
of the deformation parameter $\theta$, see \cite{Berman,Delle,openm} and below
 for details. Note also that it leads to fixed generalized noncommutativity 
parameters.
This is important to check for light-like deformations, since the 
assumption of deformation independence played an important role when these 
open brane metrics were constructed in \cite{openm}. To be more precise: 
In \cite{openm} it was assumed that a D$p$-brane solution deformed with an 
electric RR ($p-1$) form, should give a deformation independent open 
D($p-2$)-brane metric and fixed generalized noncommutativity parameter. This 
assumption was essential in the derivation of the complete open D$q$-brane 
($q=p-2$) metric and generalized noncommutativity parameter, valid for a 
one parameter deformation (see \cite{openm} for more details). The M-theory 
open membrane data were derived 
using similar assumptions. Now, since these open D-brane metrics and 
generalized noncommutativity parameters should be deformation independent for 
any one parameter deformation. It is important to show that this is indeed 
true. In \cite{openm}, this was shown for all electric and magnetic 
deformations. Below we will show that also deformations with light-like 
RR fields give deformation independent results.    

\subsection{Deforming a D$p$-brane with light-like NS-NS $B$-field}

We start by deforming a D$p$-brane with a light-like NS-NS 
B-field (i.e, $B_{+1}\neq 0$). To obtain this supergravity solution we use the 
$O(p+1,p+1)$ solution generating technique \cite{Berman,Delle,solo1}. 
Using this method gives the following solution\footnote{To be more specific, 
we use the formulas in section 2 in \cite{solo1} and deform the D$p$-brane 
with $\theta^{01}=\theta^{12}=-\theta/\sqrt{2}$. Note that we only 
include the parts of the solution which are relevant in the investigation 
below, i.e., we do not include the RR ($p+1$)-form (which, anyway, is the same 
as for an undeformed D$p$-brane).}
\begin{eqnarray}\label{NCYMLL}
ds^{2}&=&H^{-\frac{1}{2}}(2dx^{-}dx^{+}-\theta^{2}H^{-1}(dx^{+})^{2}+
(dx^{1})^{2}+(dx^{3})^{2}+\cdots +(dx^{p})^{2})\nonumber\\
& &+H^{\frac{1}{2}}(dr^{2}+r^{2}d\Omega^{2}_{8-p})\ ,\nonumber\\
e^{2\phi}&=&g^{2}H^{\frac{3-p}{2}}\ ,\quad
B_{+1}=-\theta H^{-1}\ ,\\
C_{+3\ldots p}&=&g^{-1}\theta H^{-1}\ , 
\quad H=1+\frac{Ng(\alpha')^{\frac{7-p}{2}}}{r^{7-p}}\ ,\nonumber
\end{eqnarray}
where the light-like coordinates $x^{\pm}=\frac{1}{\sqrt{2}}(x^{2}\pm x^{0})$. 
From a bound state point of view, this supergravity solution corresponds to 
 a D$p$-D($p-2$)-F1-W bound state, 
where W means that there is a wave included in the bound state. To obtain the 
supergravity duals of various noncommutative theories (NCYM) we take 
the following `magnetic' near horizon limit (similar to (5) in \cite{soloper}),
where
\begin{equation}\label{mnhl}
x^{\mu}\ ,\quad \tilde{r}=\frac{\tilde{\ell}^{2}}{\alpha'}r\ ,\quad 
\tilde{g}=g\Big(\frac{\alpha'}{\tilde{\ell}^{2}}\Big)^{\frac{p-3}{2}}\ ,\quad 
\tilde{\ell}^{2}=\theta\alpha'\ ,
\end{equation}
are kept fixed in the $\alpha'\rightarrow 0$ limit.
Inserting this limit in (\ref{NCYMLL}) gives the following supergravity 
duals
\begin{eqnarray}\label{NCYMLLn}
\frac{ds^{2}}{\alpha'}&=&\frac{1}{\tilde{\ell}^{2}}\Big(\frac{\tilde{r}}
{\tilde{R}}\Big)^{\frac{7-p}{2}}
\Big[2dx^{-}dx^{+}-\Big(\frac{\tilde{r}}{\tilde{R}}\Big)^{7-p}(dx^{+})^{2}+
(dx^{1})^{2}+(dx^{3})^{2}+\cdots +(dx^{p})^{2}\Big]\nonumber\\
& &+\frac{1}{\tilde{\ell}^{2}}\Big(\frac{\tilde{R}}{\tilde{r}}\Big)^{
\frac{7-p}{2}}(d\tilde{r}^{2}+\tilde{r}^{2}d\Omega^{2}_{8-p})\ ,\nonumber\\
e^{2\phi}&=&\tilde{g}^{2}\Big(\frac{\tilde{R}}{\tilde{r}}\Big)^
{\frac{(7-p)(3-p)}
{2}}\ ,\quad \frac{B_{+1}}{\alpha'}=-\frac{1}{\tilde{\ell}^{2}}
\Big(\frac{\tilde{r}}{\tilde{R}}\Big)^{7-p}\ ,\\
\frac{C_{+3\ldots p}}{(\alpha')^{\frac{p-1}{2}}}&=&\frac{1}{\tilde{g}
\tilde{\ell}^{p-1}}\Big(\frac{\tilde{r}}{\tilde{R}}\Big)^{7-p}\ , 
\quad \tilde{R}^{7-p}=\tilde{g}N\tilde{\ell}^{7-p}\ .\nonumber
\end{eqnarray}

Next, we calculate the open string metric, coupling constant and 
noncommutativity parameter, for a single D$p$-brane probing this background 
 solution at a distance $\tilde{r}$ from the stack. For a general background 
these are given by \cite{sw} (we use the same conventions as in 
\cite{soloper}):
\begin{eqnarray} \label{osd1}
G_{\mu\nu}&=&g_{\mu\nu}+B_{\rho\mu}g^{\rho\sigma}B_{\sigma\nu}=g_{\mu\nu}+
B^{2}_{\mu\nu}\ ,\nonumber\\
\Theta^{\mu\nu}&=&-\alpha'g^{\mu\rho}B_{\rho\sigma}G^{\sigma\nu}\ ,\\
G^{2}_{{\rm{os}}}&=&e^{\phi}\left(\frac{\det G}{\det g}\right)^{1/4}\ .
\nonumber
\end{eqnarray} 
The symmetric tensor $G_{\mu\nu}$ is the open string metric governing the
 mass-shell
condition for the open string states propagating on the probe D-brane, while
the antisymmetric tensor $\Theta^{\mu\nu}$ is
the parameter of noncommutativity between the D-brane coordinates. 
Inserting the above solution (\ref{NCYMLL}) in (\ref{osd1}), gives the 
following deformation independent result 
(before taking the limit (\ref{mnhl}))\footnote{Note that the $-$ and 1 in 
 $\Theta^{-1}$ are indices and should not be confused with the inverse of 
$\Theta$, which does not occur at all in this paper.}:
\begin{equation}\label{osd2}
G_{\mu\nu}=H^{-\frac{1}{2}}\eta_{\mu\nu}\ ,\quad
G^{2}_{\rm os}=gH^{\frac{3-p}{4}}\ ,\quad
\Theta^{-1}=\alpha'\theta\ .
\end{equation}
These open string quantities are deformation independent, 
 which means that the open string metric and coupling constant are 
independent of the deformation parameter $\theta$, and that the 
noncommutativity parameter is a constant, i.e., independent of the 
radial coordinate $r$ \cite{Berman,solo1,openm}. Together with results in 
\cite{Berman} this shows that 
all types (i.e., electric, magnetic and light-like) of one parameter 
deformations of D$p$-branes give deformation independent open string data.
Taking the magnetic near horizon limit gives the following open string data, 
and Yang-Mills coupling constant:
\begin{eqnarray}\label{osd2n}
\frac{G_{\mu\nu}}{\alpha'}&=&\frac{1}{\tilde{\ell}^{2}}\Big(\frac{\tilde{r}}
{\tilde{R}}\Big)^{\frac{7-p}{2}}\eta_{\mu\nu}\ ,\quad
G^{2}_{\rm os}=\tilde{g}\Big(\frac{\tilde{R}}{\tilde{r}}\Big)^{
\frac{(3-p)(7-p)}{4}}\ ,\nonumber\\
\Theta^{-1}&=&\tilde{\ell}^{2}\ ,\quad g^{2}_{\rm YM}=
\tilde{g}\tilde{\ell}^{p-3}\ .
\end{eqnarray}

In the solution (\ref{NCYMLL}), there is a non-zero light-like RR
$(d+1)=(p-1)$-form potential, which is dual to the light-like NS-NS $B$-field.
 This means that an open 
D$q$-brane description of the physics might be important in certain limits 
(see section 3.2.5 below). 
We therefore calculate the open D$q$-brane metric and generalized  
noncommutativity parameter ($(q+1)$-polyvector). For a one-parameter 
deformation these are given by \cite{openm}
\begin{eqnarray}\label{odd1} 
G^{{\rm od}q}_{\mu\nu}&=&\Big[1+{1\over (q+1)!}C_{q+1}^2\Big]^{{1-q\over
 1+q}}\left(g^{{\rm D}q}_{\mu\nu}+{1\over q!}(C^2_{q+1})_{\mu\nu}\right)
\ ,\nonumber\\
\Theta^{\mu_{1}\cdots \mu_{q+1}}_{\rm {od}q}&=&-(\alpha')^{\frac{q+1}{2}}
(1+\frac{1}{(q+1)!}|C_{q+1}|^{2})^{\frac{q-1}{q+1}}g^{\mu_{1}\nu_{1}}
_{{\rm D}q}C_{\nu_{1}\ldots \nu_{q+1}}G^{\nu_{2}\mu_{2}}_{\rm {od}q}
\cdots G^{\nu_{q+1}\mu_{q+1}}_{{\rm od}q},
\end{eqnarray}
where $g^{{\rm D}q}_{\mu\nu}=e^{-{2\phi\over q+1}}g_{\mu\nu}$ is the
closed D$q$-brane metric, and 
\begin{equation} \label{C2}
(C^2_{q+1})_{\mu\nu}=g^{\rho_1\sigma_1}_{{\rm D}q}\cdots g_{{\rm D}q}
^{\rho_q\sigma_q}C_{\rho_1\dots\rho_q\mu}C_{\sigma_1\dots\sigma_q\nu}\
,\quad C_{q+1}^2=g_{{\rm D}q}^{\mu\nu}(C^2_{q+1})_{\mu\nu}\ .
\end{equation} 
Inserting the above solution (\ref{NCYMLL}) in (\ref{odd1}), gives the 
following \emph{deformation independent} result
\begin{equation}\label{odqd2}
G^{{\rm od}q}_{\mu\nu}=(g^{2}H)^{-\frac{1}{q+1}}\eta_{\mu\nu}\ ,\quad
\Theta^{-3\ldots (q+2)}_{{\rm od}q}=-g(\alpha')^{\frac{q+1}{2}}\theta\ .
\end{equation}
It is important that we have obtained a deformation independent 
result, since this gives further credibility to the open D-brane 
metric and theta parameters obtained in \cite{openm}. To be more 
specific: It gives further evidence that the tensor structure of (\ref{odd1}) 
is correct, 
but unfortunately gives no information on the conformal factor, since 
$C^{2}_{q+1}=0$ for a light-like deformation. 
Taking the magnetic near horizon limit gives the following open D$q$-brane 
metric and theta parameter:
\begin{equation}\label{odqd2n}
\frac{G^{{\rm od}q}_{\mu\nu}}{\alpha'}=\frac{1}{\tilde{g}^{\frac{2}{q+1}}
\tilde{\ell}^{2}}\Big(\frac{\tilde{r}}
{\tilde{R}}\Big)^{\frac{5-q}{q+1}}\eta_{\mu\nu}\ ,\quad
\Theta^{-3\ldots (q+2)}_{{\rm od}q}=-\tilde{g}\tilde{\ell}^{q+1}\ .
\end{equation}
Note that the open D$q$-brane metric diverges in units of $\alpha'$, in the 
decoupling limit ($\hat{r}\rightarrow\infty$), while the generalized 
noncommutativity parameter is fixed and constant, which is what we expected. 

\subsection{Deforming an NS5-brane with light-like RR $C$-field}

Next we obtain NS5-brane solutions with a light-like RR ($p+1$)-form turned 
on. As we will see below, these solutions also contain a light-like RR 
($5-p$)-form potential. To obtain these solutions we S-dualize (\ref{NCYMLL}) 
for $p=5$ and then we use T-duality, using the conventions in appendix B. This 
gives the following result (we get three different solutions labeled by 1,2,3):
\begin{eqnarray}\label{NS5}
ds^{2}&=&2dx'^{-}dx'^{+}-\theta^{2}H^{-1}(dx'^{+})^{2}+
(dx'^{1})^{2}+(dx'^{3})^{2}+(dx'^{4})^{2}+(dx'^{5})^{2}
\nonumber\\
& &+H(dr'^{2}+r'^{2}d\Omega^{2}_{3})\ ,\quad 
H=1+\frac{N\alpha'}{r'^{2}}\ ,
\nonumber\\
e^{2\phi}&=&g'^{2}H\ ,\nonumber\\
&1:& p=0,4\quad C_{+}=-C_{+1345}=g'^{-1}\theta H^{-1}\ , \\
&2:& p=1,3\quad C_{+1}=C_{+345}=g'^{-1}\theta H^{-1}\ , \nonumber\\
&3:& p=2\quad C_{+15}=C_{+34}=g'^{-1}\theta H^{-1}\ , \nonumber\\
{\rm where}& &g'=g^{-1}\ , \quad x'^{\mu}=g^{-1/2}x^{\mu}\ , 
\quad r'=g^{-1/2}r\ .\nonumber
\end{eqnarray}
These three solutions correspond to the following bound states: 1. (IIA) 
NS5-D4-D0-W, 2. (IIB) NS5-D3-D1-W and 3. (IIA) NS5-D2-D2-W. Note that 2 
is S-dual to D5-D3-F1-W, 1 lifts (plus a rotation) to M5-W 
(smeared in one direction) and 3 lifts to M5-M2-M2-W (smeared in one 
direction). Next we take the following modified magnetic near horizon limit 
(similar to (20) in \cite{soloper}):
\begin{equation}\label{mnhln}
\hat{x}^{\mu}=\frac{\ell}{(\alpha')^{1/2}}x'^{\mu}\ ,\quad \hat{r}=
\frac{\ell^{3}}{(\alpha')^{3/2}}r'\ ,\quad 
\hat{g}=g'\frac{\ell^{2}}{\alpha'} ,\quad 
\ell^{2}=\theta\alpha'\ ,\quad \alpha'\rightarrow 0\ .
\end{equation}
Taking the near horizon limit for the solution (\ref{NS5}), gives
the following supergravity dual:
\begin{eqnarray}\label{NS5n}
\frac{ds^{2}}{\alpha'}&=&\frac{1}{\ell^{2}}\Big(2d\hat{x}^{-}d\hat{x}^{+}-
(d\hat{x}^{+})^{2}+\Big(\frac{\hat{r}}{\hat{R}}\Big)^{2}(d\hat{x}^{1})^{2}+
(d\hat{x}^{3})^{2}+\cdots +(d\hat{x}^{5})^{2}
\nonumber\\
& &+\Big(\frac{\hat{R}}{\hat{r}}\Big)^{2}
(d\hat{r}^{2}+\hat{r}^{2}d\Omega^{2}_{3})\Big)\ ,\quad 
\hat{R}^{2}=N\ell^{2}\ ,\nonumber\\
e^{2\phi}&=&\hat{g}^{2}\Big(\frac{\hat{R}}{\hat{r}}\Big)^{2}\ ,\nonumber\\
&1:& p=0,4\quad \frac{C_{+}}{(\alpha')^{1/2}}=\frac{1}{\hat{g}\ell}
\Big(\frac{\hat{r}}{\hat{R}}\Big)^{2}\quad \frac{C_{+1345}}{(\alpha')^{5/2}}=
-\frac{1}{\hat{g}\ell^{5}}\Big(\frac{\hat{r}}{\hat{R}}\Big)^{2}\ , \\
&2:& p=1,3\quad \frac{C_{+1}}{\alpha'}=\frac{1}{\hat{g}\ell^{2}}
\Big(\frac{\hat{r}}{\hat{R}}\Big)^{2}\quad \frac{C_{+345}}{(\alpha')^{2}}=
\frac{1}{\hat{g}\ell^{4}}\Big(\frac{\hat{r}}{\hat{R}}\Big)^{2}\ , \nonumber\\
&3:& p=2\quad \frac{C_{+15}}{(\alpha')^{3/2}}=\frac{C_{+34}}{(\alpha')^{3/2}}
=\frac{1}{\hat{g}\ell^{3}}\Big(\frac{\hat{r}}{\hat{R}}\Big)^{2}\ . \nonumber
\end{eqnarray}
These solutions are supergravity duals of Little String Theory with 
different noncommutative deformations, see section 3.

In this case it is not relevant for our investigations below to calculate open 
F1-string data since open F1-strings 
cannot end on NS5-branes, without breaking supersymmetry. However, since open 
D$p$-branes can end on NS5-branes, 
we can instead calculate the open D$p$-brane metric and generalized theta 
parameter. Inserting the above solution (\ref{NS5}) in (\ref{odd1}), gives the 
following \emph{deformation independent} 
result\footnote{In case 3 we have to use 
the open membrane data (\ref{omd}) below (since we have the special case 
with a self-dual three form), but of course use 
the closed D2-brane metric instead of the closed membrane metric, see 
\cite{openm}. Also let 
$\ell_{\rm p}^{3}\rightarrow (\alpha')^{3/2}$.}
\begin{eqnarray}\label{osd3}
G^{{\rm od}p}_{\mu\nu}&=&(g'^{2}H)^{-\frac{1}{p+1}}\eta_{\mu\nu}\ ,
\nonumber\\
&1:& p=0,4\quad \Theta^{-1345}_{{\rm od}4}=g'(\alpha')^{5/2}\theta\ ,\\
&2:& p=1,3\quad \Theta^{-1}_{{\rm od}1}=-g'\alpha'\theta\ ,\quad 
\Theta^{-345}_{{\rm od}3}=-g'(\alpha')^{2}\theta\ ,\nonumber\\
&3:& p=2\quad \Theta^{-15}_{{\rm od}2}=\Theta^{-34}_{{\rm od}2}=
-g'(\alpha')^{3/2}\theta\ .\nonumber
\end{eqnarray}
Again this is a welcomed result.
Taking the magnetic near horizon limit (\ref{mnhln}), gives the following 
open D$p$-brane metric and theta parameter:
\begin{eqnarray}\label{osd3n}
\frac{G^{{\rm od}p}_{\mu\nu}}{\alpha'}&=&\frac{1}{\hat{g}^{\frac{2}{p+1}}
\ell^{2}}\Big(\frac{\hat{r}}{\hat{R}}\Big)^{\frac{2}{p+1}}\eta_{\mu\nu}\ ,
\nonumber\\
&1:& p=0,4\quad \Theta^{-1345}_{{\rm od}4}=\hat{g}\ell^{5}\ ,\\
&2:& p=1,3\quad \Theta^{-1}_{{\rm od}1}=-\hat{g}\ell^{2}\ ,\quad 
\Theta^{-345}_{{\rm od}3}=-\hat{g}\ell^{4}\ ,\nonumber\\
&3:& p=2\quad \Theta^{-15}_{{\rm od}2}=\Theta^{-34}_{{\rm od}2}=
-\hat{g}\ell^{3}\ .\nonumber
\end{eqnarray}
Note that the open D$p$-brane metric diverges in units of $\alpha'$, in the 
decoupling limit ($\hat{r}\rightarrow\infty$), while the generalized 
noncommutativity parameters are fixed and constant, as expected. 
\subsection{Deforming an M5-brane with light-like three form $A$}

We end this section by deforming an M5-brane with a light-like three 
form $A$. This solution is easily obtained by lifting 
the solution (\ref{NCYMLL}) for $p=4$, i.e., we lift the bound state 
D4-D2-F1-W to eleven dimensions, which gives an M5-M2-M2-W bound state, where 
the M2-branes have the same charge. Lifting the solution (\ref{NCYMLL}) for 
$p=4$, to eleven dimensions gives the following bound state (we label the 
uplifting direction by $x^{5}$):
\begin{eqnarray}\label{M5LL}
ds^{2}&=&H^{-\frac{1}{3}}(2dx^{-}dx^{+}-\theta^{2}H^{-1}(dx^{+})^{2}+
(dx^{1})^{2}+(dx^{3})^{2}+(dx^{4})^{2}+(dx^{5})^{2})\nonumber\\
& &+H^{\frac{2}{3}}(dr^{2}+r^{2}d\Omega^{2}_{4})\ ,\nonumber\\
A_{+15}&=&-\theta H^{-1}\ ,\quad A_{+34}=\theta H^{-1}\ , 
\quad H=1+\frac{N\ell_{\rm p}^{3}}{r^{3}}\ ,
\end{eqnarray}
where we have used (\ref{lift}) in appendix A.
Next, by `lifting' the near horizon limit (\ref{mnhl}) we obtain the following 
eleven dimensional near horizon limit, where 
\begin{equation}\label{M5nhl}
x^{\mu}\ ,\quad \tilde{r}=\frac{\ell_{\rm m}^{3}}{\ell_{\rm p}^{3}}r\ ,\quad  
\ell_{\rm m}^{3}=\theta\ell_{\rm p}^{3}\ ,
\end{equation}
are kept fixed in the $\ell_{\rm p}\rightarrow 0$ limit. Note that this 
near horizon limit is a `tensor theory type' of limit \cite{malda12}, since 
we keep 
$r/\ell^{3}_{\rm p}$ fixed. We will also see below that this limit gives the 
supergravity dual of the (2,0) tensor theory with a light-like deformation.
Taking this near horizon limit gives the following solution 
\begin{eqnarray}\label{M5LLn}
\frac{ds^{2}}{\ell_{\rm p}^{2}}&=&\frac{1}{\ell_{\rm m}^{2}}
\Big(\frac{\tilde{r}}{\tilde{R}}\Big)
\Big[2dx^{-}dx^{+}-\Big(\frac{\tilde{r}}{\tilde{R}}\Big)^{3}(dx^{+})^{2}+
(dx^{1})^{2}+(dx^{3})^{2}+(dx^{4})^{2}+(dx^{5})^{2}\Big]\nonumber\\
& &+\frac{1}{\ell_{\rm m}^{2}}\Big(\frac{\tilde{R}}{\tilde{r}}\Big)^{2}
(d\tilde{r}^{2}+\tilde{r}^{2}d\Omega^{2}_{4})\ ,\\
\frac{A_{+15}}{\ell_{\rm p}^{3}}&=&-\frac{A_{+34}}{\ell_{\rm p}^{3}}=-\frac{1}
{\ell_{\rm m}^{3}}\Big(\frac{\tilde{r}}{\tilde{R}}\Big)^{3}\ ,
\quad \tilde{R}^{3}=N\ell^{3}_{\rm m}\ .\nonumber
\end{eqnarray}
We continue by calculating the open membrane metric and generalized 
noncommutativity parameter, which for a general background are
given by \cite{janpieter,openm,ericjanpieter}:
\begin{eqnarray}\label{omd} 
G_{\mu\nu}^{{\sss {\rm OM}}}&=&\Big(\frac{1-\sqrt{1-K^{-2}}}{K^{2}}\Big)^{1/3}
\Big(g_{\mu\nu}+\frac{1}{4}A^{2}_{\mu\nu}\Big)\ ,
\nonumber\\
\Theta^{\mu\nu\rho}_{{\sss {\rm OM}}}&=&-\ell_{\rm p}^{3}
[K(1-\sqrt{1-K^{-2}})]^{2/3}g^{\mu\mu_{1}}
A_{\mu_{1}\nu_{1}\rho_{1}}G^{\nu_{1}\nu}_{{\sss {\rm OM}}}
G^{\rho_{1}\rho}_{{\sss {\rm OM}}}
\ ,\\
K&=&\sqrt{1+\frac{1}{24}A^{2}}\ ,\quad
A^{2}_{\mu\nu}=g^{\mu_{1}\nu_{1}}g^{\mu_{2}\nu_{2}}A_{\mu_{1}\mu_{2}\mu}
A_{\nu_{1}\nu_{2}\nu}\ ,\quad A^{2}=g^{\mu\nu}A^{2}_{\mu\nu}\ .\nonumber
\end{eqnarray}
Calculating the open membrane metric and generalized noncommutativity 
parameter for (\ref{M5LL}), gives the following \emph{deformation independent}
 result:
\begin{equation}\label{omd1}
G^{{\sss {\rm OM}}}_{\mu\nu}=H^{-\frac{1}{3}}\eta_{\mu\nu}\ ,\quad
\Theta^{-15}_{{\sss {\rm OM}}}=-\Theta^{-34}_{{\sss {\rm OM}}}=
\ell_{\rm p}^{3}\theta\ .
\end{equation}
By taking the magnetic near horizon limit (\ref{M5nhl}), we obtain the 
following open membrane metric and generalized noncommutativity parameter:
\begin{equation}\label{omd2}
\frac{G^{{\sss {\rm OM}}}_{\mu\nu}}{\ell_{\rm p}^{2}}=\frac{1}
{\ell_{\rm m}^{2}}\Big(\frac{\tilde{r}}{\tilde{R}}\Big)\eta_{\mu\nu}\ ,\quad
\Theta^{-15}_{{\sss {\rm OM}}}=-\Theta^{-34}_{{\sss {\rm OM}}}=
\ell_{\rm m}^{3}\ ,
\end{equation}
which implies that we have a fixed generalized noncommutativity parameter and 
that the open membrane metric diverges in unites of $\ell_{\rm p}^{2}$, 
in the decoupling limit ($\tilde{r}\rightarrow\infty$).

We have seen in this section that for light-like deformations of 
D$p$/NS5/M5-branes, we obtain deformation 
independent open brane (F1/D$q$/M2) data in \emph{all} cases, before we 
take any near 
horizon limit. For the open D$q$-brane and open membrane data this is 
good news, since it gives further evidence that the tensor structure of 
the expressions for the open brane metrics and generalized noncommutativity 
parameters, obtained in \cite{janpieter,openm,ericjanpieter}, are 
correct. Further indications are also given by the fact 
that we in \emph{all} cases obtain open brane metrics which diverges in units 
of $\alpha'$ (or $\ell_{\rm p}^{2}$), when $\tilde{r}\rightarrow\infty$ 
(decoupling limit), after taking the near horizon limit. For 
the open string case this implies that massive open strings decouple and 
that we obtain a field theory on the probe brane. Similar for 
the open D/membrane cases we expect it to imply that there are no light open 
D/membranes in the world volume theories. 

Together with the results 
in \cite{openm} (electric and magnetic cases), the results in this section 
confirm that \emph{all} types of one parameter deformations (i.e., electric, 
magnetic and 
light-like), in all cases (i.e., deformations of D$p$/NS5/M5-branes) give 
deformation independent open brane metrics and fixed noncommutativity 
parameters.    

\section{Light-like noncommutativity and a duality web}

In this section we will discuss the various noncommutative theories, for 
which we obtained supergravity duals and relevant open F1/D$q$/M2 data in the
 last section (see also \cite{gomis,NS5}). In particular, we 
will be interested in the different phases of the theories and their 
duality relations (for IIB branes S-duality and lift to M-theory for 
IIA branes). We will also, in subsection 3.3, include a few comments  
on the relation between open membrane data in six dimensions and open string 
data in five dimensions.
\subsection{The light-like (2,0) tensor theory}
The supergravity dual (\ref{M5LLn}) corresponds to an M5-brane with a 
light-like three form $A$ turned on. The world volume theory 
is conjectured to be \cite{seiberg3,ganor4,berkooz1} the six dimensional (2,0) 
tensor theory with some kind of generalized noncommutativity structure, 
characterized by a three form $\Theta_{\sss {\rm OM}}$ \cite{ganor4}. This 
`light-like' (2,0)
 tensor theory (LL(2,0)) is obtained from the (2,0) theory by perturbing the
 (2,0) theory with a dimension 9 (${\rm mass}^{9}$) operator 
\cite{seiberg3,ganor4,berkooz1}. For a further discussion about the DLCQ 
description of this theory see \cite{seiberg3,ganor4,berkooz1}. 

In the last section we computed the open membrane metric and generalized 
noncommutativity parameter (\ref{omd2}), for the supergravity dual 
(\ref{M5LLn}). We see that we have obtained a fixed generalized 
noncommutativity parameter, which confirms the existence of the conjectured 
three form in \cite{ganor4}. However, it is not yet clear what this 
generalized noncommutativity parameter implies for the physics of 
the deformed M5-brane. It has 
been speculated \cite{others,openm,ericjanpieter}, that it gives rise to 
some kind of noncommutative loop-space structure and/or non-associative 
geometry. This three form might also be connected to a three point 
function, similar to the open string noncommutative parameter which is 
connected to the open string two-point function \cite{sw}. Also, 
from the three form generalized noncommutativity parameter we see that 
there is a length scale $\ell_{\rm m}$ in the theory where the 
`noncommutative' effects become important. This implies that for 
energies $E<<1/\ell_{\rm m}$, the (2,0) theory is a good description of 
the physics, while for energies $E\sim 1/\ell_{\rm m}$ and above, we have to 
use the LL(2,0) theory.

Equation (\ref{omd2}) implies that the open membrane metric diverges in units
of $\ell_{\rm p}^{2}$, in the decoupling limit. This also happens if 
we calculate the open membrane metric for the supergravity dual of the 
`ordinary' (2,0) theory. However, it is very different from the OM-theory 
 \cite{GMSS,us2} case. For the OM-theory supergravity dual (see e.g. 
\cite{openm}), the open 
membrane metric is instead fixed in unites of $\ell_{\rm p}^{2}$. 
This result, and the fact that the LL(2,0) theory can be obtained as a 
perturbation of the (2,0) theory, indicates that the `ordinary' (2,0) tensor 
theory and the light-like version are rather `similar' in the sense that they 
have a similar amount of degrees of freedom (d.o.f.). OM-theory instead, is 
expected to be very different compared to these tensor theories, with a much 
larger amount of d.o.f, see also discussions in \cite{gomis,berkooz1}.  
Because of this `similarity' of the two tensor theories, it would be very 
interesting to investigate if it is 
possible to obtain a map, similar to the Seiberg-Witten map \cite{sw}, for the
 LL(2,0) theory. A map of this kind might involve some kind of generalization
of the star product \cite{sw}, where the three form $\Theta_{\sss {\rm OM}}$ 
is important analogous to how the open string noncommutativity parameter 
$\Theta$ is important in the Seiberg-Witten map.  

In the next subsection we will show how the light-like three form 
$\Theta_{\sss {\rm OM}}$ is related to the five dimensional NCYM 
light-like noncommutativity parameter $\Theta$.

\subsection{Light-like theories and duality relations}

In this subsection, we continue by analyzing various non-gravitational 
theories with light-like noncommutativity in three to six dimensions, using 
the supergravity duals and open brane data, that we obtained for the deformed 
D$p$/NS5/M5-branes in the last section. 
We begin in three dimensions.

\subsubsection{Light-like noncommutativity in three dimensions}

The world volume theory of a D2-brane (probe brane) with a light-like 
$B$-field, in the decoupling limit, is a U(1) NCYM theory with 
$\Theta^{-1}=\tilde{\ell}^{2}$. The supergravity 
dual is given in (\ref{NCYMLLn}) with $p=2$. The divergence of the open 
string metric (in units of $\alpha'$) implies that massive open string 
modes decouple. This theory has a dimension-full Yang-Mills coupling constant
 $g^{2}_{\rm YM}=\tilde{g}\tilde{\ell}^{-1}$, which means, as usual, that the 
theory has to be completed at low energies $E\leq g^{2}_{\rm YM}$. At low 
energies, we expect the light-like NCYM theory to be completed by an
SO(8) invariant super conformal M2-brane theory (the bound state D2-F1-D0-W 
lifts (plus a rotation) to M2-W, where the wave is expected to be irrelevant 
at low energies).   

\subsubsection{Light-like noncommutativity in four dimensions}

Next, in four dimensions (see also \cite{gomis,NS5,cai6,paolo}), we have a 
D3-brane 
(probe brane) with a light-like $B$-field (dual to a light-like 
$C_{2}$-field). The supergravity dual is given by (\ref{NCYMLL}), 
with $p=3$, and the open string data is given in (\ref{osd2n}). The open 
string metric diverges, in unites of $\alpha'$, in the decoupling limit, 
while the noncommutativity parameter is fixed \mbox{$\Theta^{-1}=
\tilde{\ell}^{2}$}.
 This implies that we have obtained a U(1) noncommutative Yang-Mills theory, 
with light-like noncommutativity and dimensionless Yang-Mills coupling 
constant $g^{2}_{\rm YM}=\tilde{g}$. This theory is unitary and 
renormalizable, which has been shown in \cite{gomis}. It was 
also argued that the strong coupling limit of this theory is 
another NCYM theory with light-like noncommutativity (see also 
\cite{NS5,paolo}). 
The relations between the two theories coupling constants and 
noncommutativity parameters are given by
\begin{equation}
g^{2}_{{\rm YM}({\rm s})}=\frac{1}{g^{2}_{\rm YM}}\ ,\quad 
\Theta_{({\rm s})}^{-3}=-g^{2}_{\rm YM}\Theta^{-1}\ ,
\end{equation}
where the index (s) means the S-dual parameter. To obtain this result, we 
S-dualize the supergravity dual, calculate the 
open string data, and compare the open string data for the two S-dual 
supergravity duals. Note that this is rather different compared to 
NCYM with space-space noncommutativity, which is S-dual to NCOS in 
four dimensions \cite{gopa}. 

The above NCYM theory with non-zero $\Theta^{-1}$, can be generalized to 
light-like NCYM with both $\Theta^{-1}$ and $\Theta^{-3}$ non-zero. The 
S-dual of this theory is given by a light-like NCYM theory which also has 
non-zero $\Theta^{-1}_{({\rm s})}$ and $\Theta^{-3}_{({\rm s})}$. If 
$\Theta^{-1}=\Theta^{-3}=\tilde{\ell}^{2}$, the parameters of
 the two (S-dual) theories are related as follows:
$g^{2}_{{\rm YM}({\rm s})}=\frac{1}{g^{2}_{\rm YM}}$ and 
$\Theta_{({\rm s})}^{-1}=-\Theta_{({\rm s})}^{-3}=g^{2}_{\rm YM}
\tilde{\ell}^{2}$.

\subsubsection{Light-like noncommutativity in five dimensions}

In five dimensions we have a D4-brane with a light-like $B$-field (dual to a 
light-like $C_{3}$-field). In this case we also have a U(1) NCYM theory with 
$\Theta^{-1}=\tilde{\ell}^{2}$, but with a dimension-full coupling constant 
$g^{2}_{\rm YM}=\tilde{g}\tilde{\ell}$. This implies that the theory 
is not renormalizable, i.e., the theory breaks down above energies 
$E\sim (g^{2}_{\rm YM})^{-1}$. This can 
be seen if we introduce an effective dimensionless 
coupling constant $g^{2}=E g^{2}_{\rm YM}$, where $E$ is the energy scale. 
The effective coupling constant $g^{2}$
is only much less then one for energies $E<<1/g^{2}_{\rm YM}$. For larger 
energies this theory can be completed by the world volume theory of an 
M5-brane with a light-like three form $A$-field
 turned on, wrapped on a circle parallel to the M5-brane. This world 
volume theory is the light-like (2,0) theory with generalized noncommutativity 
parameter $\Theta^{-15}_{\sss {\rm OM}}=-\Theta^{-34}_{\sss {\rm OM}}$, see 
section 3.1. Wrapping the M5-brane on a 
circle with radius $R$ (in rescaled units), gives the following relations 
between the parameters 
of the two theories
\begin{equation}\label{32}
g^{2}_{\rm YM}=R\ ,\quad \Theta^{-1}=\frac{\Theta^{-15}_{{\sss {\rm OM}}}}{R}\
 .
\end{equation}
These relations are obtained by comparing the 
two supergravity duals (\ref{NCYMLLn}) with $p=4$, and (\ref{M5LLn}), 
using the results in Appendix A.
Equation (\ref{32}) implies that we have the following two `phase 
diagrams'\footnote{What we mean here by phase diagram is that we start at 
low energy and let the energy $E$ become larger and larger and at the 
same time we give the description (theory) which is relevant at a certain 
energy. At certain energies the description of the physics has to be 
changed, since new effects appear (e.g. noncommutativity). Note that 
strictly speaking a transition from one description to another does not 
have to be a phase transition. For example, going from SYM to NCYM is not a 
`true' phase transition.} for 
a D4-brane with a light-like $B$-field turned on, after taking the 
decoupling limit (see also subsection 3.3 for additional comments on this 
case):

{\bf 1}. 
$\Theta_{\sss {\rm OM}}^{1/3}<<\Theta^{1/2}$: phase 1. for 
energies $E<<1/\Theta^{1/2}$ the physics on the probe brane can
 be described by `ordinary' five dimensional super Yang-Mills, since the 
noncommutativity is negligible; phase 2. for energies above $1/\Theta^{1/2}$
but much less then $1/g^{2}_{\rm YM}=1/R$, light-like NCYM provides a good 
description; phase 3. for energies $E\sim 1/g^{2}_{\rm YM}=1/R$ and 
above, we have to use the light-like (2,0) tensor theory on a circle with 
radius $R$.

{\bf 2}. 
$\Theta_{\sss {\rm OM}}^{1/3}>>\Theta^{1/2}$: Start with `ordinary' SYM for 
energies 
$E<<1/R$, while for energies $E\sim 1/R$ we have to use the (2,0) tensor 
theory on circle. For energies $E\sim 1/\Theta_{\sss {\rm OM}}^{1/3}$ and 
above we use the light-like (2,0) tensor theory on a circle. Note that there 
is no NCYM phase in this case. 

\subsubsection{Light-like noncommutativity in six dimensions}

{\bf Type IIB D5/NS5-branes}: 

In six dimensions, the world volume theory for a D5-brane with 
light-like $B$-field (dual to a RR $C_{4}$-field), is NCYM theory with 
noncommutativity parameter $\Theta^{-1}=\tilde{\ell}^{2}$ and 
dimension-full coupling constant $g^{2}_{\rm YM}=\tilde{g}\tilde{\ell}^{2}$ 
(the supergravity dual is given in (\ref{NCYMLLn}) with $p=5$). This 
theory is not renormalizable, and has to be completed for energies above 
$E\sim 1/g_{\rm YM}$, which is seen from the effective coupling constant 
$g^{2}=g^{2}_{\rm YM}E^{2}$. The six dimensional light-like NCYM theory can be 
completed by (1,1) Little String theory
with light-like noncommutativity (light-like LST). To obtain the relevant 
phase diagram we use the supergravity duals (\ref{NCYMLLn}) with $p=5$, and 
(\ref{NS5n}) case 2, which 
are S-dual to each other\footnote{The S-dual of the IIB D5-brane is the IIB 
NS5-brane. For an NS5-brane without any RR-fields turned on, the physics 
(after decoupling gravity) 
at energies $E\sim T^{1/2}_{\rm s}$ and above, is described by the 
`ordinary' Little String theory \cite{seiberg11,aharony11}, i.e., (1,1) LST 
theory for the IIB NS5-brane 
and (2,0) LST theory for the IIA NS5-brane. For energies $E<<T^{1/2}_{\rm s}$ 
the effective theory is (1,1) six dimensional SYM for IIB and a (2,0) tensor 
theory for IIA. For a DLCQ description of these theories, see 
\cite{berkooz}.}. We have to use the S-dual 
`picture' for energies above $E\sim 1/g_{\rm YM}$, since light-like NCYM 
becomes strongly coupled. The relations between the parameters of the two 
supergravity duals, as well as the relations between the relevant open brane
quantities are given by
\begin{equation}\label{para1}
\hat{g}=\frac{1}{\tilde{g}}\ ,\quad \ell^{2}=\tilde{g}\tilde{\ell}^{2}
\quad \Theta^{-1}_{{\rm od}1}=-\Theta^{-1}\ .  
\end{equation}
The tension of the little strings on the NS5-brane is given by 
\begin{equation}\label{para2a}
T_{\rm s}=\frac{1}{g^{2}_{\rm YM}}=\frac{1}{\tilde{g}\tilde{\ell}^{2}}=
\frac{1}{\ell^{2}}\ ,
\end{equation}
where we have used (\ref{para1}) and $g^{2}_{\rm YM}$ is the six dimensional
Yang-Mills coupling constant. Note that we in this paper ignore factors 
of $2\pi$ in the definition of tension. Before we obtain the phase diagrams, 
we note 
that for the deformed NS5-brane we have both a two form RR field as 
well as a four form RR field, in the supergravity dual. These 
RR fields are not independent but are dual to each other. This duality 
is due to the self-duality of the M-theory three form on the M5-brane. 
The RR two form leads to a noncommutativity parameter $\Theta_{{\rm od}1}$, 
while the four form leads to a generalized noncommutativity parameter 
$\Theta_{{\rm od}3}$, see (\ref{osd3n}) case 2.
Note also that the length scales where these noncommutativity parameters 
become important are different. This implies that they are important at 
different energy scales. However, we expect both to be important at very high 
energies (i.e., energies $E$ larger then both $1/\Theta^{1/2}_{{\rm od}1}$ and 
$1/\Theta^{1/4}_{{\rm od}3}$). In \cite{berkooz,gomis} a possible DLCQ 
description of this light-like LST were discussed. However only 
the RR two form were considered. The exact role of the RR four form and 
$\Theta_{{\rm od}3}$ are unclear to us at the moment, but we expect them to 
play some part in a complete description of the above light-like LST.
Using (\ref{para1}) and (\ref{para2a}), 
we obtain the following two phase diagrams:

{\bf 1}. 
$\Theta^{1/2}<<g_{\rm YM}$. If we start at low energy $E<<1/g_{\rm YM}$,
 the D5-brane (probe brane) physics can be described by `ordinary' six 
dimensional SYM. At energies $E\sim 1/g_{\rm YM}=T_{\rm s}^{1/2}$ we have to 
use the S-dual description (D5$ \rightarrow$ NS5). This means that we can use 
`ordinary' (1,1) Little String theory \cite{seiberg11} as long as 
$E<<1/(\Theta_{{\rm od}3})^{1/4}$. Above this energy we
 have to use light-like (1,1) LST theory where $\Theta^{-345}_{{\rm od}3}=
\hat{g}\ell^{4}$ is important. For energies 
$E\sim 1/(\Theta_{{\rm od}1})^{1/2}$ and above also the noncommutativity 
parameter $\Theta^{-1}_{{\rm od}1}=\hat{g}\ell^{2}$ becomes important. 

{\bf 2}. 
$\Theta^{1/2}>>g_{\rm YM}$. We start at low energy 
$E<<1/(\Theta)^{1/2}$, where we have `ordinary' six dimensional 
SYM. For energies above this but much less then $1/g_{\rm YM}$ we have 
light-like NCYM. At energies $E\sim 1/g_{\rm YM}$ we switch to the S-dual 
description, which implies that  we have light-like (1,1) LST theory where 
both $\Theta_{{\rm od}3}$ and $\Theta_{{\rm od}1}$ are relevant. 

\vspace{0.2cm}
{\bf Type IIA NS5-brane}: 

Next, we consider the type IIA 
NS5-brane with light-like RR three 
form turned on (the supergravity dual is given by (\ref{NS5n}) case 3). A 
light-like three form means that we have (2,0) LST theory 
with generalized noncommutativity parameter $\Theta^{-15}_{{\rm od}2}=
\Theta^{-34}_{{\rm od}2}=-\hat{g}\ell^{3}$. 

The type IIA NS5-brane with light-like RR three form, can be interpreted as an
 M-theory M5-brane with light-like three form turned on, with a small 
transverse circle 
with radius $R_{\rm T}$ (in rescaled units). The 
relations between the relevant open D2/M2 data, are given by
\begin{equation}\label{para2}
\ell_{\rm m}^{3}=\hat{g}\ell^{3}\ ,
\quad \Theta_{{\rm od}2}=\Theta_{\sss {\rm OM}}\ ,\quad T_{\rm s}=
\frac{R_{\rm T}}{\ell^{3}_{\rm m}}=\frac{1}{\ell^{2}}\ .
\end{equation}
These relations implies that we have the following two phase diagrams for the 
 type IIA NS5-brane with light-like three form, after taking the 
decoupling limit:

{\bf 1}. 
$R_{\rm T}<<\ell_{\rm m}<<\ell$. At low energy $E<<T^{1/2}_{\rm s}$ we 
have the (2,0) NS5 tensor theory as usual. For energies 
$E\sim T^{1/2}_{\rm s}$ but much less then $1/\Theta_{\rm od2}^{1/3}$, 
the physics on the probe brane is described by the `ordinary' LST, while 
for energies $E\sim 1/\Theta^{1/3}_{\rm od2}$ and above we have to use the 
light-like LST with $\Theta_{\rm od2}^{-15}=\Theta_{\rm od2}^{-34}$. 

{\bf 2}. 
$R_{\rm T}>>\ell_{\rm m}>>\ell$. Start with the (2,0) M5 tensor theory with 
a transverse circle with radius $R$,
for energies $E<<1/\Theta_{\sss {\rm OM}}^{1/3}=1/\ell_{\rm m}$. Then for 
energies above this but much less then $T^{1/2}_{\rm s}$, we use the light-like
 (2,0) tensor theory, while for energies $E\sim T^{1/2}_{\rm s}$ and above, 
we use the light-like LST with 
$\Theta_{\rm od2}^{-15}=\Theta_{\rm od2}^{-34}$.

For a discussion about a possible DLCQ 
description of this light-like LST, see \cite{berkooz,gomis}.

For the IIA NS5-brane with light-like RR five form (dual to a RR one form) 
turned on we have light-like LST with generalized noncommutativity 
parameter $\Theta^{-1345}_{\rm od4}=\hat{g}\ell^{4}$ (the supergravity dual is
 given by (\ref{NS5n}) case 1). We do not 
expect the dual one form to lead to any noncommutativity. The two phase 
diagrams are as follows:

{\bf 1}. 
$\Theta_{\rm od4}^{1/5}<<T^{-1/2}_{\rm s}$. We start at low energies 
$E<<T^{1/2}_{\rm s}$, where we have the `ordinary' (2,0) NS5 tensor 
theory. At $E\sim T^{1/2}_{\rm s}$ we use `ordinary' LST, while for 
energies $E\sim 1/\Theta_{\rm od4}^{1/5}$ and above, we have to use the 
light-like LST with generalized noncommutativity parameter 
$\Theta_{\rm od4}^{-1345}$.   

{\bf 2}. 
$\Theta_{\rm od4}^{1/5}>>T^{-1/2}_{\rm s}$. At low energies $E<<T^{1/2}$ we 
have the (2,0) M5 tensor theory with 
a transverse circle with radius $R=\hat{g}\ell$, while at energies 
$E\sim T^{1/2}$ and above, 
we have the light-like LST with $\Theta_{\rm od4}^{-1345}$.

\subsubsection{Light-like noncommutativity and generalized gauge theories}

Before we end this subsection, we would like to make a few comments on a 
possible dual description of the D$p$-branes with light-like $B$-field. As 
we have seen before, using open strings to define the D$p$-branes led to 
NCYM theories with light-like noncommutativity. On a D$p$-brane with 
light-like $B$-field there is also a dual light-like ($p-1$)-form RR 
$C$-field, as can be seen from the supergravity duals (\ref{NCYMLLn}). This 
implies that there, in principle, is a dual formulation of the D$p$-brane 
with open D$q$-branes ($q=p-2$) ending on the D$p$-brane. From (\ref{odqd2n}) 
we see that the open D$q$-brane metric diverges, in units of $\alpha'$, in the
decoupling limit, and that there is a fixed generalized noncommutativity 
parameter $\Theta_{{\rm od}q}^{-3\ldots (q+2)}\neq 0$. This indicates that 
we can describe the physics on the probe D$p$-brane with some kind of 
generalized gauge theory with generalized noncommutativity parameter 
$\Theta_{{\rm od}q}$. For $q=1$ we have `ordinary' noncommutativity, while 
for $q>1$, there is some generalization of noncommutativity. We expect that 
these theories, if they indeed exists, can be formulated in a similar way as 
the generalized gauge theories D$q$-GT, defined in \cite{soloper} 
(which are obtained by deforming a D$p$-brane with a magnetic RR $(p-1)$ 
form). Defining them in a similar way as in \cite{soloper}, gives a gauge 
coupling which is the same as the Yang-Mills coupling $g^{2}_{\rm YM}$ and a 
fixed generalized noncommutativity parameter $\Theta_{{\rm od}q}$, see 
(\ref{odqd2n}). This implies that these generalized gauge theories, similar to
 NCYM, only are complete theories for $p=3$ ($q=1$). 

It is unclear to us if these generalized gauge theories are important or not. 
For example, they can not be used to `complete' NCYM for $p\neq 3$. 
Studying them further might, however, give 
important information on the generalized noncommutative structure they seem 
to inhibit. Since the relevance of these theories are not clear to us and 
the fact that we have a well understood NCYM description in these 
cases, we have chosen to not include them in the phase diagrams for the 
deformed D$p$-branes above.

\subsection{Relations between open string/membrane data}
In this subsection we will make a few comments on the relations between open 
string data in five dimensions and open membrane data in six dimensions. 
As we mentioned before, in \cite{janpieter,openm,ericjanpieter} the complete 
open membrane metric was obtained, as well as an open 
membrane generalized noncommutativity parameter \cite{openm,ericjanpieter}. 
 Further, in \cite{openm,ericjanpieter} it was 
shown that if one tries to relate the open string metric and coupling 
constant in five dimensions, to the open membrane metric in six dimensions, 
there are a few subtleties. In particular, 
if one reduces the open membrane metric on an `electric' circle, 
then the open membrane metric reduces to the open string metric and coupling 
constant (see \cite{openm,ericjanpieter} for details). With 
electric reduction we mean the following: Assume the M-theory three form 
$A$ has nonzero $A_{012}$ and $A_{345}$ components if we use an 
SO(1,5)/SO(1,2)$\times$SO(3) parametrization (see \cite{openm} for details) 
then an electric reduction implies that the reduction is in the $x^{2}$ 
direction, i.e., $A_{012}\rightarrow B_{01}$. Similarly, a reduction in e.g. 
the $x^{5}$ direction is called a magnetic reduction.
If one instead reduce on a `magnetic circle', 
then it does not seem possible to obtain the open string metric and coupling 
constant. However, in \cite{openm} it was shown that the open membrane metric
reduces to an open D2-brane metric (also obtained in \cite{openm} with  
another method as well) in the magnetic case. These results
 are in fact natural considering that in the electric case the electric 
component of three form reduces to an electric component of the NS-NS 
$B$-field, while in the magnetic case it reduces to an electric component 
of the RR three form. For a general rank four reduction, 
it was shown in \cite{ericjanpieter} that it is not possible to 
obtain the open string metric and coupling constant from the open membrane 
metric, which also seems natural since the rank 4 reduction is a mixture of 
an `electric' \emph{and} a `magnetic' reduction. For a rank 4 reduction it 
also does not seem to be possible to obtain an open D2-brane metric. 
It would be very interesting to further study the rank 4 case, since it is 
unclear what the failure 
of the reduction implies for what kind of description one should use on the 
D4-brane, in this case. It is possible that both open strings 
and open D2-branes are important for the description of the world volume 
physics on the D4-brane in this case. 

In this paper, we are in particular interested in how the open string data 
are related to the open membrane data if the M-theory three form is 
light-like on the M5-brane and reduces to a light-like two form $B$ 
on the D4-brane (rank 2). This case was not investigated in 
\cite{openm,ericjanpieter}. For the light-like case we know that $A^{2}$,
 which is defined in (\ref{omd}) is $A^{2}=0$ (implies $B^{2}=B^{4}=0$). 
 Next, if we use the reduction formulas derived in \cite{openm}, 
we obtain that the open 
membrane metric and generalized noncommutativity parameter \emph{can} be
 related to the open string metric, coupling constant and noncommutativity 
parameter, as follows:
\begin{equation}\label{omsd}
\frac{G_{\mu\nu}^{\sss {\rm OM}}}{\ell^{2}_{\rm p}}=(G_{\rm os}^{2})^{-
\frac{2}{3}}
\frac{G_{\mu\nu}}{\alpha'}\ ,\quad \frac{G_{yy}^{\sss {\rm OM}}}{\ell^{2}_{\rm
 p}}=\frac{(G_{\rm os}^{2})^{\frac{4}{3}}}{R^{2}}\ ,\quad \frac{
\Theta^{\mu\nu y}_{\sss {\rm OM}}}{R}=\Theta^{\mu\nu}\ ,
\end{equation}
where $\mu,\nu$ are the five dimensional indices and $y$ is the direction in 
which we reduce. As expected the open brane quantities (\ref{omsd})
are related analogously to how the closed brane quantities (see (\ref{lift}) 
below) are related.
This implies that we can reduce the open membrane data to open string data, 
in the light-like case. The open D2-brane metric can also be obtained with a 
similar argument. The crucial fact in that these reductions work, is that 
$A^{2}=0$, because this implies that the conformal factor in the open membrane
 metric (\ref{omd}) is equal to 1. The results are also further confirmed by 
comparing the open string/membrane data in (\ref{osd2}) $p=4$, and 
(\ref{omd1}), using $\ell_{\rm p}^{2}=g^{2/3}\alpha'$ and 
$R^{2}=g^{2}\alpha'$. 

\section{NCYM with space-space noncommutativity and strong coupling}

In this section\footnote{To obtain the results in this section we have 
benefited from discussions with M. Cederwall, U. Gran, B.E.W. Nilsson and 
P. Sundell.}, we will obtain the theory which is the 
strong coupling (high energy) limit of five dimensional NCYM with 
$\Theta^{12}=\Theta^{34}\neq 0$. 
We begin by showing that the light-like (2,0) theory is not the strong 
coupling (high energy) limit of NCYM with $\Theta^{12}=
\Theta^{34}\neq 0$. Naively one might think that the two theories are 
connected, since the bound state D4-D2-D2-D0, which is used to obtain the NCYM 
supergravity dual, lifts to an M5-M2-M2-W bound state. However, as we will 
see below, the supergravity solutions are in fact \emph{not} related, once we 
take the near horizon limit. The relevant part of the supergravity solution, 
corresponding to 
the bound state D4-D2-D2-D0 (with equal D2-brane charge), is given 
by\footnote{We use the O$(p+1,p+1)$ method in \cite{Delle,Berman,solo1}. This 
bound state have been obtained earlier in \cite{oz}, using other methods. Note 
that we have not included the RR five form.}
\begin{eqnarray}\label{D4220}
ds^{2}&=&H^{-\frac{1}{2}}\Big(-(dx^{0})^{2}+\frac{1}{h}((dx^{1})^{2}+\cdots +
(dx^{4})^{2})\Big)+H^{\frac{1}{2}}(dr^{2}+r^{2}d\Omega^{2}_{4})\ ,
\nonumber \\
e^{2\phi}&=&g^{2}H^{-\frac{1}{2}}h^{-2}\ ,\quad B_{12}=B_{34}=gC_{012}=
gC_{034}=-\theta' (Hh)^{-1}\ ,\\
gC_{0}&=&\theta'^{2}H^{-1}\ ,\quad h=1+\theta'^{2}H^{-1}\ .\nonumber
\end{eqnarray}
If we continue by taking the near horizon limit (\ref{mnhl}) (with $\theta'$ 
instead of $\theta$), we obtain the 
supergravity dual of five dimensional NCYM with Yang-Mills coupling constant 
and noncommutativity parameter given by
\begin{equation}
g^{2}_{\rm YM}=\tilde{g}\tilde{\ell}\ ,\quad \Theta^{12}=\Theta^{34}=
\tilde{\ell}^{2}\ .
\end{equation} 
For energies above $E\sim 1/g^{2}_{\rm YM}$ this NCYM theory is strongly 
coupled and we have to change description. We therefore lift the type IIA 
solution to M-theory, in order to find a theory which 
`completes' the above NCYM theory.
Lifting the type IIA D4-D2-D2-D0 bound state to M-theory gives the 
following M5-M2-M2-MW bound state\footnote{We use the relations in  
appendix A, as well as rewriting the solution in light-like coordinates. 
This bound state solution has earlier been obtained, using different methods, 
in \cite{NS5,eric1,roy3,ulf1}.} 
\begin{eqnarray}\label{M5LLA}
ds^{2}&=&(Hh)^{-\frac{1}{3}}[2dx^{-}dx^{+}-2\theta'^{2}(Hh)^{-1}(dx^{+})^{2}+
(dx^{1})^{2}+\cdots +(dx^{4})^{2}]\nonumber\\ 
& &+(Hh)^{\frac{2}{3}}(dr^{2}+r^{2}d\Omega^{2}_{4})\ ,\quad x^{\pm}=
\frac{1}{\sqrt{2}}(x^{5}\pm x^{0})\ ,\\
A_{+12}&=&A_{+34}=-\sqrt{2}\theta' (Hh)^{-1}\ , 
\quad h=1+\theta'^{2}H^{-1}\ ,\quad H=1+\frac{N\ell_{\rm p}^{3}}
{r^{3}}\ .\nonumber
\end{eqnarray}
Comparing this solution with (\ref{M5LL}) in section 2, we see that 
(\ref{M5LLA}) also
 correspond to an M5-M2-M2-MW bound state, since \mbox{$H'=Hh=1+
\theta'^{2}+\frac{N\ell_{\rm p}^{3}}{r^{3}}$} is a harmonic function. 
Note that the two solutions are very similar, except that 
the constants in the harmonic function are different. This difference in the 
harmonic functions is, however, very important when we take the near 
horizon limit (\ref{M5nhl}), because the constant in the harmonic function 
$H'$ has a $\theta'$ dependence, where $\theta'$ scales (\ref{M5nhl}). 
 This implies that after we have taken the near horizon limit (\ref{M5nhl}) 
in (\ref{M5LL}) and (\ref{M5LLA}), 
we obtain two \emph{different} results. This naturally also leads to 
different results for the open membrane metric. Looking closer, we find that 
 in the form the solution (\ref{M5LLA}) is written, the limit 
(\ref{M5nhl}) is `trivial'. To be more precise: taking the limit (\ref{M5nhl}) 
for the solution (\ref{M5LLA}), gives a solution which still is a 
\emph{complete} 
M5-M2-M2-MW solution (let $\ell_{\rm m}\rightarrow \ell_{\rm p}$, which gives
 (\ref{M5LLA}) with $\theta'=1$, and $H'=Hh=1+\frac{N\ell_{\rm p}^{3}}
{r^{3}}$). 
The conclusion of this is that the supergravity dual of five dimensional NCYM 
with rank 4 space-space noncommutativity 
and the supergravity dual (\ref{M5LLn}) of the light-like (2,0) theory 
are not connected. This can also be seen by reducing 
the supergravity dual (\ref{M5LLn}), in a direction $x^{2}$ ($x^{2}=\frac{1}
{\sqrt{2}}(x^{+}+x^{-})$), which gives a type IIA supergravity solution 
which has a singularity in the metric at finite $\tilde{r}$. This  
solution is clearly not the same as (\ref{D4220}) after one has taken the near
 horizon limit (\ref{mnhl}). For related results concerning the M5-M2-M2-MW 
bound state and near horizon limits, see \cite{ulf1}. 

We still have not answered the question about the strong coupling limit of 
five dimensional NCYM with $\Theta^{12}=\Theta^{34}\neq 0$. It is clear 
from the above discussion, that it can not be the six dimensional light-like 
(2,0) theory. 

Next, the ten dimensional supergravity dual lifts to an 
eleven dimensional supergravity solution, which approaches flat 
eleven dimensional space-time, in the $\tilde{r}\rightarrow\infty$ limit. 
In the $\tilde{r}\rightarrow\infty$ limit we get the 
following metric (after rescaling the coordinates)
\begin{equation}
ds^{2}=-(dx^{0})^{2}-dx^{0}dx^{5}+(dx_{9})^{2}\ ,
\end{equation}
where $dx_{9}$ is the line element of nine dimensional flat space and $x^{5}$ 
is the `uplifted' direction such that $x^{5}\sim x^{5}+2\pi R$.  
The result we get here is similar to 
what happens if one lifts the NS5-D0 bound state followed by taking the 
near horizon limit, see \cite{GMSS}. The NS5-D0 bound state
 is used to obtain the supergravity dual of the OD0 theory \cite{GMSS,NS5}
(a six dimensional theory containing light D0-branes), which can be obtained
 as a DLCQ compactification of M-theory with $n$ units of DLCQ momentum 
in the presence of a
transverse M5-brane. If we compare this to our case we see that there are 
some similarities. For example, in the D4-D2-D2-D0 bound state there is a 
D0-brane. As in the NS5-D0 case this D0-brane is light in the decoupling limit,
 with tension (mass) given by 
$T_{\sss {\rm D0}}=\frac{1}{\tilde{g}\tilde{\ell}}$.
Note that $T_{\sss {\rm D0}}=1/g^{2}_{\rm YM}$. We expect that these light 
D0-branes can be found as noncommutative solitons 
(see e.g. \cite{gross1} for an introduction to noncommutative solitons)
in the five dimensional NCYM theory. It would be interesting to 
explicitly obtain this noncommutative soliton solution in the NCYM theory.

With the above results and 
comparing with the NS5-D0 case \cite{GMSS}, we make the following observation.
 Since the five dimensional NCYM theory with $\Theta^{12}=\Theta^{34}=\Theta$ 
is only a low energy effective theory we propose that the world
 volume theory of the above deformed D4-brane at high energies (i.e., 
energies $E\sim 1/g^{2}_{\rm YM}$ and above) is given by a 
DLCQ compactification of M-theory in the presence of a `parallel' 
M5-brane (i.e., one of the M5-brane directions is the compactified $x^{5}$ 
direction) and two transverse M2-branes. The relations between the NCYM data 
($g_{\rm YM}^{2}$ and $\Theta$), the  
compactification radius $R$ and the effective M-theory Planck mass 
(in rescaled units) are given by:
\begin{equation}\label{rel4}
R=g^{2}_{\rm YM}\ ,\quad M^{3}_{\rm eff}=\frac{1}{g^{2}_{\rm YM}\Theta}\ .
\end{equation}
This gives the following two phase diagrams: 

{\bf 1}. 
$g^{2}_{\rm YM}<<\Theta^{1/2}$. Start with `ordinary' 
super-Yang-Mills theory up to energies $E<<1/\Theta^{1/2}$, followed by 
NCYM with $\Theta^{12}=\Theta^{34}=\Theta$ for energies
$E<<1/R=1/g^{2}_{\rm YM}$. At energies $E\sim 1/R$ we have to lift to 
M-theory (i.e., use the above DLCQ compactification of M-theory), since 
the NCYM theory is strongly coupled.

{\bf 2}.
$g^{2}_{\rm YM}>>\Theta^{1/2}$. This case is the same as case 1 except that 
there is \emph{no} NCYM phase.


\vspace{0.4cm}
\Large \textbf{Acknowledgments}
\vspace{0.2cm}
\normalsize

We are grateful to R. Argurio, M. Cederwall, U. Gran, M. Nielsen, B.E.W. 
Nilsson and P. Sundell for valuable discussions and comments.

\appendix

\section{Conventions for S-duality, T-duality and lift to eleven dimensions}

In this appendix we give the conventions we have used for S-duality, T-duality
 and lift of type IIA solutions to eleven dimensions.

The S-duality rules for a type IIB solution (in the string frame) are given
 by (in the case of zero axion):
\bea \label{S}
d\tilde{s}^2&=& e^{-\phi}ds^2\ ,\quad e^{\tilde{\phi}}=e^{-\phi}\ ,
\nonumber\\
\tilde{B}&=&C_{2}\ ,\quad \tilde{C}_{2}=-B\ ,\\
\tilde{C}_{4}&=&C_{4}+B\wedge C_{2}\ .\nonumber
\eea

For uplifting of a type IIA solution to eleven dimensions we use the
following relations:
\begin{eqnarray} \label{lift}
{ds_{11}^2\over \ell_p^2}&=&e^{-2\phi/3}{ds_{IIA}^2\over
\alpha'}+e^{4\phi/3}({dx^{11}\over R}-{C_1\over \sqrt{\alpha'}})^2\ ,
\nonumber\\
{A_3\over\ell_p^3}&=&{C_3\over (\alpha')^{3/2}}+{dx^{11}\over R}\wedge 
{B_2\over \alpha'}\ ,
\end{eqnarray}
where $x^{11}$ has radius $R$ and $\ell_p$ is the eleven-dimensional
Planck length, which are given by $R=g\sqrt{\alpha'}$ and
$\ell_p=g^{1/3}\sqrt{\alpha'}$, where $g$ is the asymptotic value
of the dilaton.

For T-duality between IIA and IIB solutions we use the Buscher rules
\cite{bucher} (in the string frame)

\begin{eqnarray}\label{TNS}
\tilde{g}_{yy}&=&\frac{1}{g_{yy}}\ ,\quad \tilde{g}_{\mu\nu}=g_{\mu\nu}-
\frac{g_{\mu y}g_{\nu y}-B_{\mu y}B_{\nu y}}{g_{yy}}\ ,\nonumber\\
\tilde{B}_{\mu y}&=&\frac{g_{\mu y}}{g_{yy}}\ ,\quad \tilde{B}_{\mu\nu}=
B_{\mu\nu}-\frac{B_{\mu y}g_{\nu y}-g_{\mu y}B_{\nu y}}{g_{yy}}\ ,\\
\tilde{g}_{\mu y}&=&\frac{B_{\mu y}}{g_{yy}}\ , \quad e^{2\tilde{\phi}}=
\frac{e^{2\phi}}{g_{yy}}\ , \nonumber
\end{eqnarray}
where $y$ denotes the Killing coordinate with respect to which the
T-dualization is applied, while $\mu$,$\nu$ denote any coordinate direction
other then $y$. The RR $p$-form fields transforms as \cite{sei3}:
\begin{eqnarray}\label{TRR}
\tilde{C}_{(p)\mu\cdots \nu\rho y}&=&C_{(p-1)\mu\cdots \nu\rho}-
(p-1)\frac{C_{(p-1)[\mu\cdots\nu\mid y\mid}g_{\rho]y}}{g_{yy}}\ ,\nonumber\\
\tilde{C}_{(p)\mu\cdots \nu\rho\sigma}&=&C_{(p+1)\mu\cdots \nu\rho\sigma y}+
pC_{(p-1)[\mu\cdots\nu\rho}B_{\sigma]y}\\
& & +p(p-1)\frac{C_{(p-1)[\mu\cdots\nu\mid y\mid}B_{\rho\mid y\mid}
g_{\sigma] y}}{g_{yy}}\ .\nonumber
\end{eqnarray}

\section{`Flat space scaling' an example}
In this appendix we give an example, which shows the equivalence 
between the `super gravity dual' approach used in this paper (see also e.g. 
\cite{Berman,soloper,openm}) to noncommutative theories 
and the `flat space scaling' approach (see e.g. \cite{sw,gomis,gopa,GMSS}). 
With the `flat space scaling' approach
we mean that the noncommutative theory is obtained as a limit of 
string/M-theory, in flat space. As an example, 
we will deform the M5-brane with a light-like three form $A$. The 
decoupling limit in this case is given by:
\begin{eqnarray}\label{m5LL}
\epsilon&\rightarrow& 0\ ,\quad \ell_{\rm p}^{2}=\ell_{\rm m}^{2}\epsilon\ ,
\quad 
A_{+15}=-A_{+34}=-\epsilon^{-3/2}\ ,\nonumber\\
g_{++}&=&-\epsilon^{-3}\ ,\quad g_{-+}=g_{+-}=1\ ,
\quad x^{\pm}=\frac{1}{\sqrt{2}}(x^{2}\pm x^{0})\ ,\\
g_{ab}&=&\delta_{ab}\ ,\quad a,b=1,3,4,5\ ,\quad 
g_{mn}=\epsilon^{3}\delta_{mn}\ ,\quad m,n=6,\ldots,10\ .
 \nonumber
\end{eqnarray} 
This gives the following (fixed) open membrane metric and generalized 
noncommutativity parameter:
\begin{equation}\label{omd4}
G^{{\sss {\rm OM}}}_{\mu\nu}=\eta_{\mu\nu}\ ,\quad \Theta_{\sss 
{\rm OM}}^{-15}=-\Theta_{\sss {\rm OM}}^{-34}=\ell_{\rm m}^{3}\ .
\end{equation}
If we now compare (\ref{m5LL}) and (\ref{omd4}) with the supergravity dual 
(\ref{M5LLn}) and open membrane data in (\ref{omd2}), we 
have a perfect match between the two different approaches if we identify 
\begin{equation}
\Big(\frac{\tilde{r}}{\tilde{R}}\Big)=\epsilon^{-1}\ . 
\end{equation}
This implies that 
the two approaches give the same decoupling limit. Conclusion: taking the 
decoupling limit (\ref{m5LL}) above corresponds in the supergravity 
approach, to first obtaining the M5-M2-M2-W bound state (\ref{M5LL}), then 
taking the near horizon limit (\ref{M5nhl}), which gives the supergravity 
dual (\ref{M5LLn}), followed by taking the 
$\Big(\frac{\tilde{r}}{\tilde{R}}\Big)=\epsilon^{-1}\rightarrow\infty$ 
(decoupling) limit. 

To obtain the other flat space scaling limits, which corresponds to the 
supergravity duals (\ref{NCYMLLn}) and (\ref{NS5n}), are straight-forward, 
and we will therefore not include them here.

\end{document}